\documentstyle[art12]{article}

\begin{document}
\begin{titlepage}
\begin{flushright}
INP MSU Preprint-95-12/376\\
QMW-PH-95-13\\
hep-ph/yymmnn\\
17-May-1995
\end{flushright}
\begin{center}
{\bf\boldmath{
{On a Method of Identification of Leptoquarks \\ Produced  in
$ep$ Collisions.}
}}
\end{center}
\bigskip
\begin{center}
{V. Ilyin, A. Pukhov, V. Savrin, A. Semenov, \\
{\small\it Institute of Nuclear Physics of Moscow State University,
        119899 Moscow, Russia}\\
 and \\
 W. von Schlippe \\
{\small\it Queen Mary \& Westfield College, London, England}
 }
\end{center}\bigskip
\vfill
\begin{abstract}
We analyse numerically manifestations of the radiative
amplitude zero (RAZ) effect in single leptoquark production
associated with
hard photon emission. We present some quantitative conclusions on the
possibility to distinguish leptoquark charges produced in $ep$ collisions
taking account of three-body final state subprocesses and of
proton structure
functions. Applying this method to HERA and possible LEP+LHC
experiments we
show that the RAZ analysis
can serve as a tool to determine the leptoquark electric charge
up to large leptoquark masses.
\end{abstract}

\vfill
\begin{center}
{(To be submitted to Physics Letters)}
\end{center}
\end{titlepage}

\section{Introduction.}

In paper \cite{paper1} we have proposed a method for
distinguishing
the charges of leptoquarks\footnote{The problem of
identification of LQ is discussed now in the literature, see
recent publication \cite{Mike} where the determination of the leptoquark
properties in polarized $e\gamma$ collisions is considered.}
 (LQ) produced in the reaction
$$ e^\pm+p\rightarrow \gamma+LQ+X.$$
The idea is based on the radiative amplitude zero (RAZ) effect,
i.e. the absence of photon emission at some directions depending on
the leptoquark electric charges. In this paper we give a detailed
numerical
analysis of this method applying it to HERA and possible
LEP+LHC experiments.

In Ref. \cite{paper1} we have derived the analytical formulas for the
unpolarized cross sections of the hard subprocesses
\begin{equation}
e^\pm+q\rightarrow \gamma+LQ
\label{eq:iii}
\end{equation}
We have shown that they are
proportional\footnote{For the vector leptoquark this is
correct only in the case of Yang-Mills coupling.}
to a factor $(Q_eu-Q_qt)^2$. So, at some photon emission
angles where this factor vanishes the cross sections have  exact zeros
whose positions 
are completely determined by the electron
(positron) and quark charges, $Q_e$ and $Q_q$, respectively.

Of course in reality the leptoquarks decay into two fermions.
Considering the electron decay channel we have got
the following clear signature: the hard photon, the electron
(or positron)
and a hard (quark) jet with complete kinematical reconstruction of the
subprocess:
\begin{equation}
e^\pm+q\rightarrow \gamma+e^\pm+q
\label{eq:iiii}
\end{equation}

In this paper we analyse numerically the manifestations of the RAZ
effect
for the case of this three-body final state subprocess. We make some
quantitative conclusions on a possibility to distinguish the charges
of leptoquarks produced in $ep$ collisions taking
account of the proton structure functions.

Our calculations were made with the help of the CompHEP package
\cite{CompHEP}
created for automatic calculation of cross sections at tree level
in the Standard Model and
beyond. For the numerical integration over three-particle
phase space and for event generation we have used the BASES package
\cite{BASES}.

\section{Calculation framework \label{calcframe}}

The amplitude of subprocess $e^-+d \to e^-+d+\gamma$ is represented by
the Feynman diagrams displayed in Figs. 1,\, 2.
In Fig. 1
the diagrams
corresponding to the signal LQ production process (2) are
represented. The diagrams of Fig. 2
correspond to the SM contribution: deep
inelastic scattering associated with hard photon emission.

Using the CompHEP package we have
calculated analytically the unpolarized squared matrix
elements  for
subprocesses (\ref{eq:iiii}). But in contrast to our previous paper
\cite{paper1}
we have introduced finite widths into the LQ propagators. Then we
have evaluated numerically the contribution
of separate (quark) constituents to the
integrated cross sections by the convolution of the subprocess cross
section with the corresponding parton distribution function:
\begin{equation}
\sigma(s)\;=\; \int^1_{x_{min}} dx\; q(x,Q^2)\, \int
    d\Phi_3
   \frac{d\hat\sigma}{d\Phi_3}
   \Theta_{cuts}(E_{\gamma},\vartheta_\gamma, ...).
                                               \label{eq:iiiii}
\end{equation}

Here $d\Phi_3$ is the element of 3-body phase space;
$\hat\sigma$ is the cross section of
subprocess (\ref{eq:iiii});
$s$ is the squared CMS energy of the electron-proton
system; the squared CMS energy of the hard subprocess
equals $\hat s = x s$;
the quark distribution function is denoted by $q(x,Q^2)$
and the 4-momentum transfer scale is taken to be $Q = \sqrt{\hat s}$.
The photon emission angle is denoted by $\vartheta_\gamma$. The
direction $\vartheta_\gamma=0$ is along the proton beam. As the lower
 bound
$x_{min}$ we use the value ${M^2}/{s}$ to take into account the
quasi-resonant peak, where $M$ is the LQ mass. The function
$\Theta_{cuts}(E_{\gamma},\vartheta_\gamma, \ldots)$ corresponds to
the kinematical
cuts which we impose to get a realistic distribution and values of
cross sections. In particular this function has to include cuts
on the photon
energy and on the angles between the photon and all charged particles
in (2) to remove infrared and collinear divergences. It is also
necessary
to apply cuts for a reliable separation of the hard jet.

All the fermion masses are put equal to zero in
our calculations.

The numerical analysis of the corresponding cross sections
is given for two cases:
\begin{itemize}
 \item HERA: $\sqrt{s} = 296\,\mbox{GeV}$,
    electron beam energy ${\cal E}_e = 26.7\,\mbox{GeV}$,
      integrated luminosity $L = 100\,\mbox{pb}^{-1}$;

  \item LEP+LHC: $\sqrt{s}=1740\,\mbox{GeV}$, ${\cal E}_e = 100$ GeV,
     annual integrated luminosity  $L = 1000\,\mbox{pb}^{-1}$.
\end{itemize}

One of the general restrictions on the LQ-fermion interaction
is the chirality of the fermion-LQ
coupling (see \cite{BRW} and references therein).
Therefore we consider only either couplings
with left-handed leptons or with right-handed ones.
In both cases we use
in the numerical analysis  an ``electroweak'' value for this
constant, $\lambda = 0.3$.

As to LQ mass we analyse cross sections in the following ranges
\begin{itemize}
 \item  HERA: \hspace{1cm} $150\,\mbox{GeV} < M < 290\,\mbox{GeV}$;
 \item  LEP+LHC: \hspace{1cm} $200\,\mbox{GeV} < M < 1500\,\mbox{GeV}$.
\end{itemize}

 For the parton densities
we used the parametrizations STEQ2p \cite{CTEQ} and MRS-A \cite{MRS},
which take account of recent HERA data.
Both parametrizations gave the same results within calculation error.

Let us describe in more detail the cuts which we apply in
the laboratory frame.

 For a proper background analysis we introduce kinematical cuts for this
reaction in such a way as to suppress contributions from
the Standard Model Feynman
diagrams (see Fig. 2). Recall that for the  problem under consideration
the LQ mass
is supposed to be already known from resonant LQ production. So
we can introduce a narrow cut on the invariant mass of the
outgoing electron (or positron) and the (quark) jet:
\begin{equation}
 |(p^{out}_e + p^{out}_q)^2 \; -\; M^2|\; <\; 6 M \Gamma,
                                       \label{eq:s45cut}
\end{equation}
where $\Gamma$ is the LQ width. We have calculated $\Gamma$ at tree
level for each value of the LQ mass.

We introduce also a cut on the angle between the outgoing electron
(positron) and the quark: $\vartheta_{e q} > 10^\circ$.  In the case of
LQ decay the products are moving  back-to-back in the LQ rest frame,
so this cut suppresses the standard model contributions and does
not change the LQ signal.

Then we have to introduce cuts on the energies and emission angles
of the outgoing electron (positron) and quark to exclude
the forward and backward cones and various soft contributions:
\begin{equation}
   E_e,E_q\, >\, 10\; \mbox{GeV}, \qquad
       10^\circ<\vartheta_e,\vartheta_q<170^\circ.
                                                    \label{eq:eq-cuts}
\end{equation}

We also apply cuts on the photon energy and emission angle:
$E_\gamma\,>\, 1\mbox{ GeV}$,
$10^\circ \,<\, \vartheta_\gamma \,<\, 170^\circ$.

Finally we introduce cuts on the angles between the outgoing photon
and the
electron (positron) and quark:
$\vartheta_{\gamma e} > 10^\circ, \;\;\;
  \vartheta_{\gamma q} > 10^\circ$.
These cuts remove the collinear divergences.

For this set of cuts and with $x_{min}=(200\,\mbox{GeV})^2/s$
the cross sections of reaction (\ref{eq:iiii})
calculated in the Standard Model are:

\begin{tabular}{ll}
   HERA & \hspace{1cm} $\sigma_{\mbox{\small SM}}\sim\,
   0.5\,\mbox{fb}$; \\
   LEP+LHC & \hspace{1cm} $\sigma_{\mbox{\small SM}}\sim\,
   40\,\mbox{fb}$.
\end{tabular}

So, for our analysis we have an SM background of
less than 1 event for HERA (assuming an integrated luminosity of
$100\,\mbox{pb}^{-1}$) and less than 50 events per year for LEP+LHC.
 For
larger LQ masses from the corresponding ranges (see above) we introduce
larger cuts $x_{min}$, so the SM background is smaller.

The final results presented in this paper were calculated for reaction
(\ref{eq:iiii})
with a complete set of tree level diagrams, i.e. taking into account
 the
SM diagrams, the contributions of the signal reaction (\ref{eq:iii})
and the SM-LQ interference terms.

\section{Numerical results and RAZ analysis}

In this section we present results of the numerical analysis for the
left-handed sector of  LQ-fermion interactions.
For the right-handed sector the results for all channels are the same
as for the corresponding processes in the left-handed sector due to
identical analytical formul\ae\  \cite{paper1}.
For the leptoquarks we use the notation of \cite{BRW} (see also the
detailed table of LQ quantum numbers in \cite{paper1}).

First we note that the cross sections of interest are of order
$ 1\, \mbox{pb}$ for HERA and $ 10\, \mbox{pb}$ for LEP+LHC.
These values are obtained for reaction (\ref{eq:iiii})
with the kinematical cuts discussed in section \ref{calcframe}
and with $M=200\,\mbox{GeV}$ for HERA and $M=300\,\mbox{GeV}$ for
LEP+LHC with an electroweak value of the fermion-LQ coupling constant
$\lambda=0.3$. So, there are about one-hundred events at HERA
and more than ten thousand events at LEP+LHC. This must be compared
with the SM background, which is less than 1 event at HERA and 50
events per year at LEP+LHC.

A typical example of a distribution in $x$
is shown in Fig. 3,
for the $S^{-1}_3$ leptoquark with
$\sigma=0.82\,\mbox{pb}$ for HERA at $M=200\,\mbox{GeV}$
and $\sigma=13\,\mbox{pb}$ for LEP+LHC at $M=300\,\mbox{GeV}$.
One can see that there is a sharp peak near
$x_{min}={M^2/s}$.

The cross sections decrease with increasing leptoquark mass.
This is shown in Fig. 4
for HERA and in Fig. 5
for LEP+LHC.

In Fig. 6
we present the angular distributions
of hard photons for the case of scalar leptoquarks and
$u$ and $d$ valence quarks.
Again we present these distributions at $M_{LQ}=200\,\mbox{GeV}$
for HERA and at $M_{LQ}=300\,\mbox{GeV}$ for LEP+LHC.

For HERA we took 16 bins, $10^\circ$ per bin. Due to the small
statistics
(even for an integrated luminosity of
$100\,\mbox{pb}^{-1}$) a finer binning does not allow us to recognize
the RAZ effect. For LEP+LHC we took 40 bins, $4^\circ$ per bin.
We see explicit RAZ for
$S_3^{-1}$ and $R_2^{-{1\over 2}}$ leptoquarks at different values of
RAZ angles in good agreement with Ref. \cite{paper1} where the
RAZ angles
were estimated analytically.
For the $e^- u$ channel (leptoquarks $S_1,\, S^0_3$) this
angular distribution has no radiative amplitude
zero\footnote{The drop in the bin
$130^\circ\,<\,\vartheta_\gamma\,<\,140^\circ$
in this channel is a statistical fluctuation.}
and the distribution in the $e^+ d$ case is similar.

The angular distributions in $\vartheta_\gamma$ for  vector leptoquarks
with Yang-Mills couplings and valence quarks in the initial
state have a behaviour similar to the corresponding cases for scalar LQ;
they are therefore not shown.

Also for channels with sea quarks the distributions in the photon
emission
angle are similar but with
a significantly smaller level of cross sections
(see Fig. 5
).
Note that for HERA there is no
statistics for any of the processes in the channels with sea quarks.

Considering these distributions we see
that for HERA the RAZ analysis seems to be
statistically unsupported in the case of leptoquark masses
greater than $200\,\mbox{GeV}$ (for $\lambda=0.3$).
However from the HERA data analysis corresponding to an integrated
luminosity of $\approx 425 \,\mbox{nb}^{-1}$ \cite{HERA}
for the $R_2$ and $\tilde R_2$ scalar LQ and
for the $U_1$, $\tilde U_1$ and $U_3$ vector LQ the established
mass limit
is smaller than $200\,\mbox{GeV}$ (for $\lambda=0.3$).
In this small mass range between the experimental limit and
$200\, \mbox{GeV}$  the determination of
the LQ electric charge is still possible by the proposed method
if this new boson is discovered at HERA.

For LEP+LHC there is enough statistics for a reliable
RAZ analysis up to large leptoquark
masses. Consider, as a criterion, 10 events per $4^\circ$ bin, in which
case we can say with confidence whether RAZ is present or absent.
There are 40 bins in the interval
$10^\circ < \vartheta_\gamma < 170^\circ$, so the lower
limit for the cross section is $0.4\,\mbox{pb}$ to be certain
that RAZ is
statistically visible.  From the leptoquark mass dependence
of the cross
sections (see Fig. 4
and Fig. 5)
we derive the upper limits for leptoquark masses
with visible RAZ. These mass
limits are shown in Table \ref{tab:RAZ-masses}.

We see that for LEP+LHC the RAZ analysis,
and indeed a measurement of the leptoquark charge,
can be made in the $e^- p$ and $e^+ p$ channels separately up
to $M\sim 500-600\,\mbox{GeV}$ for all
types of leptoquarks. For higher masses, up to $850-900\,\mbox{GeV}$,
all types of leptoquarks could also be
identified but only some of them in the $e^- p$
channel and others in the $e^+ p$ channel.
The upper limit for the RAZ analysis
is about $M\sim1\,\mbox{TeV}$, see details in
Table \ref{tab:RAZ-masses}.
For the vector case the situation is slightly better
for $U_3$ and $\tilde U_1$ leptoquarks. Here the RAZ effect can be
reliably determined up to $M\sim 1200\,\mbox{GeV}$.

Note that the pairs ($S_1$, $S_3^0$) and ($U_1$, $U_3^0$)
(in the left-handed sector) will remain unresolved by
this method, because of their equal charges and third
components of isospin.

Finally we note that our analysis was made for the reaction with the
electron decay mode of LQ. So only leptoquarks with the third
components of
isospin equal to $-1$ and $-{1\over 2}$ from the corresponding
isotriplets
and isodoublets were
considered in the analysis (with the only exception of $R_2$ and
$V_2$ in the right-handed sector). To analyse leptoquarks with
complementary
isospin projections from the same multiplet it is necessary
to investigate
the neutrino decay mode of LQ. However the signature of these processes
includes missing energy due to the neutrino emission.
 So one has to carry
out the analysis differently in this case. This analysis will
be presented elsewhere.

\section*{Conclusions}
Our principal conclusion is that for some types of leptoquarks
(in the left-handed sector: $S_3^{-1}$ and $R_2^{-{1\over 2}}$ scalars,
$V_2^{-{1\over 2}}$ and $U_3^{-1}$ vectors;
in the right-handed sector: $\tilde S_1$ and $R_2^{-{1\over 2}}$
 scalars,
$V_2^{-{1\over 2}}$ and $\tilde U_1$ vectors)
the {\it radiative amplitude zero} is present
in the distribution in the photon emission angle.
This effect can be detected at LEP+LHC up to large values of
leptoquark masses: 900 GeV for $S_3^{-1}$ and $V_2^{-{1\over2}}$,
1 TeV for $R_2^{-{1\over2}}$ and 1.2 TeV for $U_3^{-1}$ and $\tilde U_1$.

The RAZ analysis (whether the radiative amplitude zero
 is present or not)
gives us a possibility to determine the
leptoquark electric charge (and, as a result, its other quantum numbers).
This method is available up to rather large leptoquark masses (up to
1 TeV at LEP+LHC).
Even at HERA there still exists a small mass region (near 200 GeV)
for some leptoquarks where the RAZ effect can be observed.

 For vector leptoquarks the exact RAZ effect is present only
 in the case of
Yang-Mills type of photon-LQ coupling. So we have got a fairly sensitive
tool to measure the anomalous magnetic moment $\kappa$
in the case of RAZ. A detailed analysis of this possibility will be
presented in a future publication.

All our numerical results were obtained for the electroweak value of the
fermion-LQ coupling constant, $\lambda=0.3$. For other values
of $\lambda$
numerical estimates can be obtained by rescaling our results
(cross sections have $\lambda^2$ as a factor).

\section*{Acknowlegements}

This work was partly supported by the Royal Society of London as
a joint project between Queen Mary \& Westfield College (London) and
Institute of Nuclear Physics, Moscow State University, and
by INTAS (project 93-1180).
V.I. and V.S. wish to thank QMW for their hospitality and
the possibility to work during their visit in London.
WvS wishes to thank INP-MSU for hospitality.
The work of V.I., A.P. and A.S. was partly supported by Grants
M9B000 and M9B300 from the International Science Foundation.

\section*{Table}

\begin{table}[hp]
\begin{center}
\begin{tabular}{||c|c|cccc|cccc||}
\hline
\multicolumn{2}{||c|}{\raisebox{0ex}[3ex][1ex]{$\ell q$ channel}}
           &$e^-u$&$e^-d$&$e^-\bar u$&$e^-\bar d$&$e^+u$
           &$e^+d$&$e^+\bar u$&$e^+\bar d$ \\ \hline \hline
\raisebox{0ex}[3ex][1ex]{scalar LQ}& L &$S_1$,$S_3^0$&$S_3^{-1}$
&$R_2^{-{1\over 2}}$&$\tilde R_2^{-{1\over 2}}$
             &$R_2^{-{1\over 2}}$&$\tilde R_2^{-{1\over 2}}$
&$S_1$,$S_3^0$&$S_3^{-1}$ \\
         & R & $S_1$ & $\tilde S_1$ & $R_2^{-{1\over 2}}$
& $R_2^{1\over 2}$
             & $R_2^{-{1\over 2}}$ & $R_2^{1\over 2}$ & $S_1$
& $\tilde S_1$ \\ \hline
\multicolumn{2}{||c|}{$\mbox{max}\;M$ {\small [GeV]}}
    & 900 & 900 & 540 & 610 & 980 & 830 & 670 & 500 \\ \hline \hline
\raisebox{0ex}[3ex][1ex]{vector LQ}& L &$\tilde V_2^{-{1\over 2}}$
&$V_2^{-{1\over 2}}$&$U_3^{-1}$
             &$U_1$,$U_3^0$&$U_3^{-1}$&$U_1$,$U_3^0$
&$\tilde V_2^{-{1\over 2}}$
             &$V_2^{-{1\over 2}}$ \\
         & R & $V_2^{1\over 2}$ & $V_2^{-{1\over 2}}$ & $\tilde U_1$
& $U_1$
         &$\tilde U_1$ & $U_1$ & $V_2^{1\over 2}$
&$V_2^{-{1\over 2}}$ \\ \hline
\multicolumn{2}{||c|}{$\mbox{max}\;M$ {\small [GeV]}}
       &1140 & 910 & 610 & 690 & 1180 & 830 & 660 & 660 \\ \hline
\end{tabular}
\end{center}
\caption{LEP+LHC. The RAZ upper limits for LQ masses
 (in the positron channels anti-LQ are produced). L-row corresponds to
the interaction of LQ with left-handed leptons and R-row to
right-handed ones.}
\label{tab:RAZ-masses}
\end{table}

\section*{Figures captions}

\begin{itemize}
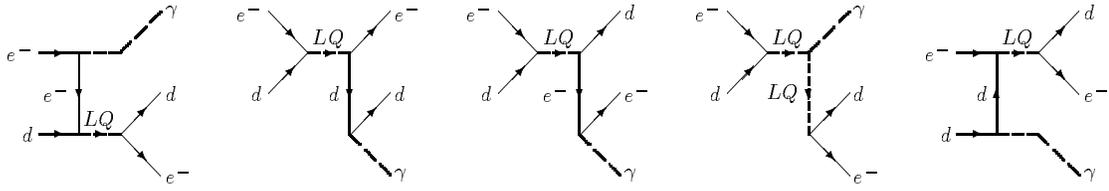
\item[Figure 1:] Feynman diagrams with LQ signal.

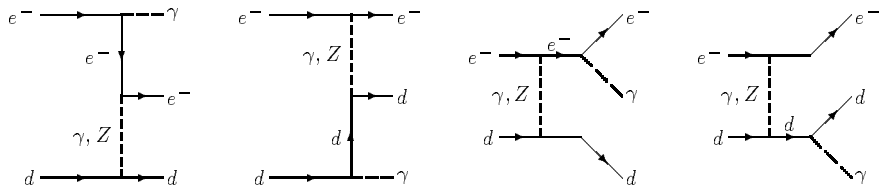
\item[Figure 2:] Feynman diagrams for subprocesses $e^-+q\rightarrow
e^-+q+\gamma$ in SM.

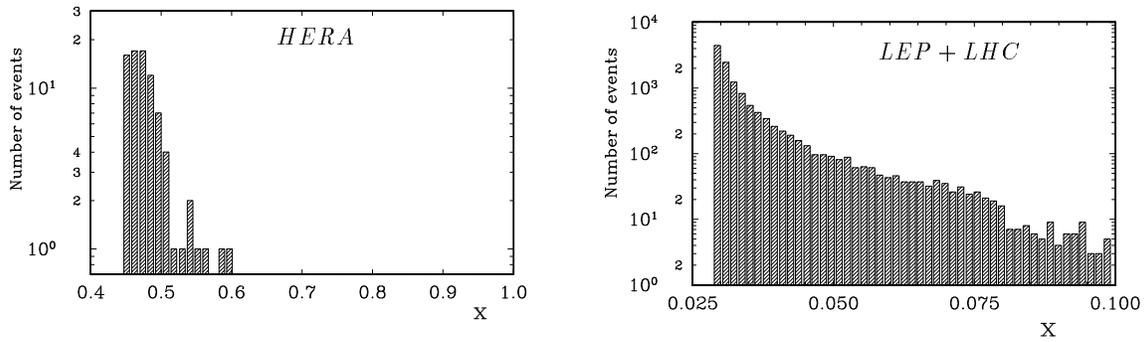
\item[Figure 3:] Distributions in $x$ for $e^-+d\rightarrow\gamma
+e^-+d$ with the $S^{-1}_3$ leptoquark contribution.
   These distributions are obtained at $M=200\,GeV$ for HERA
   ($\sigma=0.82\,\mbox{pb}$)
        and at $M=300\,GeV$ for LEP+LHC ($\sigma=13\,\mbox{pb}$)
        with kinematical cuts applied.

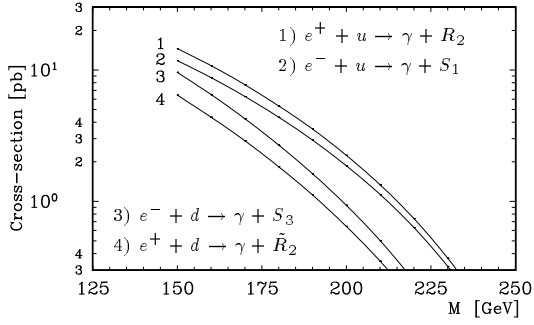
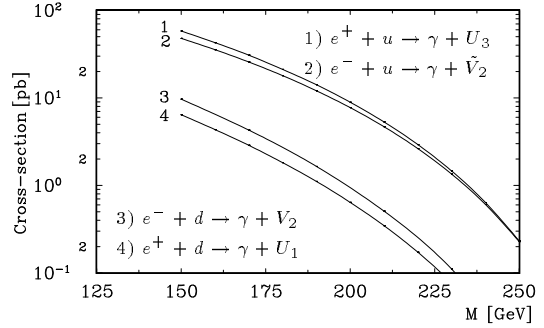
\item[Figure 4:] Cross sections in dependence on the LQ mass at HERA.

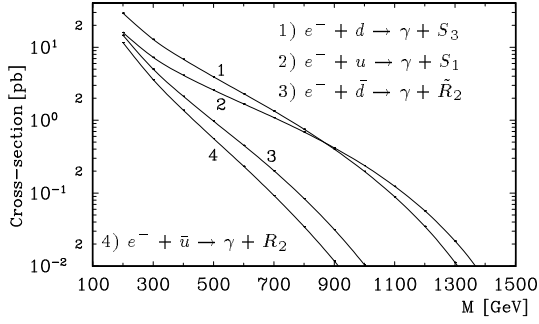
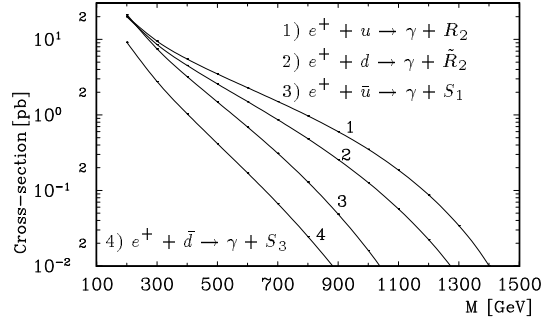
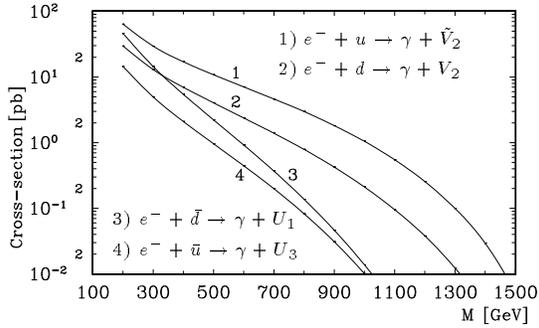
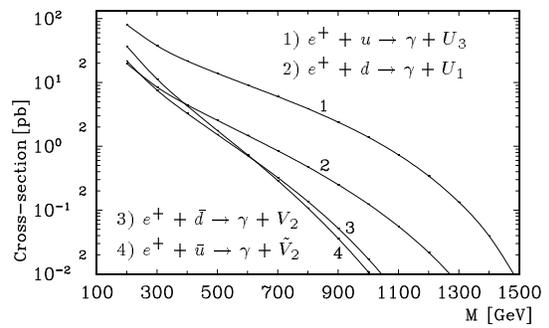
\item[Figure 5:] Cross sections in dependence on the LQ mass
                 at LEP+LHC.

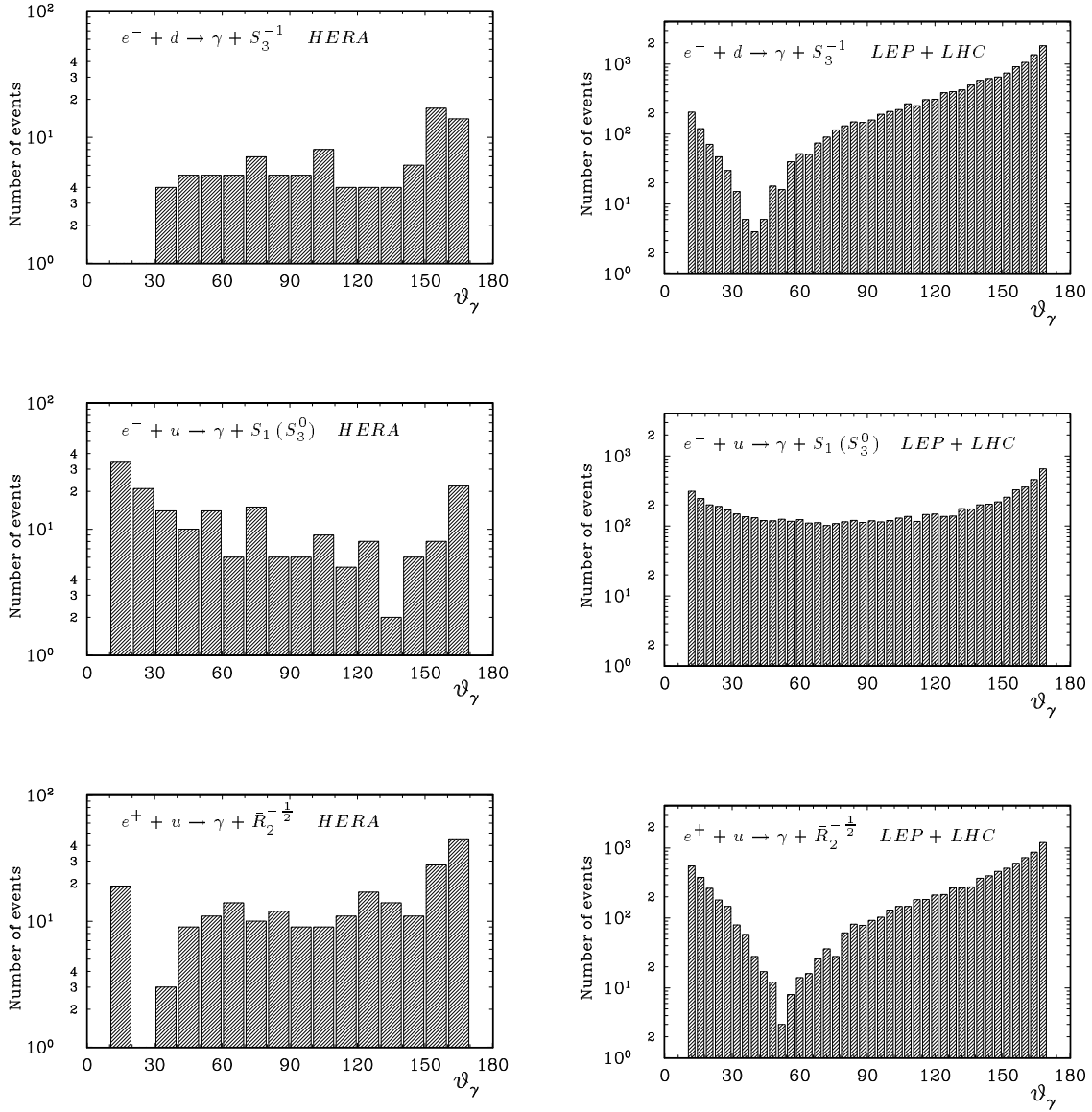
\item[Figure 6:] Distributions in the photon emission angle
    $\vartheta_\gamma$
   for valence quarks ($M=200\,\mbox{GeV}$ for HERA
   and $M=300\,\mbox{GeV}$ for LEP+LHC) in the case of scalar LQ.

\end{itemize}

\end{document}